\documentclass[a4paper,12pt]{article}

\usepackage{graphicx,epstopdf}

\usepackage{amsmath}

\usepackage{hyperref}
\hypersetup{colorlinks=true,linkcolor=blue,citecolor=blue}

\usepackage{subfigure}

\usepackage{geometry}
\begin{document}

% **************************************

\noindent
\textbf{\large
	Accurate potential energy curve for helium dimer retrieved from
	viscosity coefficient data at very low temperatures
	}

\vspace{1cm}

\noindent
\textbf{\small \'Ederson D'M. Costa$^{\rm a}$, Nelson H. T. Lemes$^{\rm a*}$
	and Jo\~ao P. Braga$^{\rm b}$}

\noindent
$^{\rm a}$Instituto de Qu\'{\i}mica, Universidade Federal de Alfenas,
Rua Gabriel Monteiro da Silva, 700 - Centro, Alfenas/MG, 37130-000, Brazil.

\noindent
$^{\rm b}$Departamento de Qu\'{\i}mica, Universidade Federal de Minas Gerais,
Av. Ant\^onio Carlos, 6627 - Pampulha, Belo Horizonte/MG, 31270-010, Brazil.

\vspace{1cm}
\noindent

\vspace{\stretch{1}}

\noindent{*nelson.lemes@unifal-mg.edu.br}

\thispagestyle{empty}

\newpage

% **************************************

\section*{Abstract}

The long range potential of helium-helium interaction, which requires accurate
\textit{ab initio} calculation, due to the small value of the potential depth,
approximately 11 K ($9.5\times 10^{-4}$~eV~= 0.091~kJ/mol) at 2.96~{\AA},
will be obtained in this study by an alternative technique.
This work presents a robust and consistent procedure that provides the
long range potential directly from experimental data. However, it is difficult
to obtain experimental data containing information regarding such a small potential
depth. Thereby, sensitivity analysis will be used to circumvent this difficulty,
from which viscosity data at lower temperatures ($<5$~K) were chosen as appropriate
data to be used to retrieve the potential function between 3 and 4~{\AA}. The linear
relationship between the potential energy function and the viscosity coefficient
will be established under quantum assumptions and the Bose-Einstein statistic.
The use of quantum theory is essential, since the temperatures are below 5~K.
The potential obtained in this study describes the viscosity with an average
error of 1.68~\% that is less than the experimental error (5~\%), with the
results being similar to those obtained for recent \textit{ab initio} potentials.

\hspace{1cm}

\noindent
\textbf{Keywords:} Variable phase method. Viscosity coefficient.
Low temperature. Sensitivity analysis.

\thispagestyle{empty}

\newpage

% **************************************

\section{Introduction}

Potential energy functions play a central role in chemistry,
and from these functions, the properties of a system can,
in principle, be determined. Often, the potential energy function
is obtained from \textit{ab initio} methods, but it can also be
determined from experimental data using inverse techniques.
Inverse problems theory to handle experimental data has been
applied to refine potential energy functions from the
second virial coefficient,\cite{Cox1980,Lemes2006}
differential cross-sections\cite{Ho1989,Lemes2008}
and phase shifts.\cite{Lemes2014}

The helium diatomic system at low temperatures has received
considerable attention in recent years due to quantum and
relativistic effects in their thermodynamics and
transport properties.\cite{Cencek2012} The theoretical interpretation
of the results has been performed with the
\textit{ab initio} potential function.\cite{Costa2013}
The helium dimer has a very small potential well depth,
measuring approximately 11 K,\cite{Varandas2010,Przybytek2010}
consequently, highly accurate calculations are required
to obtain the potential energy function.\cite{Varandas2010,Przybytek2010}

The present paper presents an alternative way to refine the
helium potential energy function from viscosity coefficient data
at low temperatures. Few studies have been conducted along
this line and often at high temperatures,\cite{Maitland1981}
limit in which classical theory is valid, and in parametric form,
usually for a Lennard-Jones potential energy curve.\cite{Kim2014}

To gain insight into this problem, an investigation into the
sensitivity\cite{Lemes2006,Ho1989} of the viscosity coefficient
to the potential energy function is conducted, and an adequate
temperature range for experimental data was observed.
The present study shows a higher sensitivity for long range
potential in the temperature range below 5~K. Since it is necessary
to use viscosity coefficient data at low temperatures
in an inverse procedure, a quantum strategy to refine the
potential energy curve is necessary and will be presented in this study.

The inverse problem was divided into two parts: in the first part,
the cross-section is obtained from the viscosity coefficient,
whereas in the second part, the potential energy function is determined from
the cross-section. The solution to the first part was determined using
the Tikhonov regularization under a Laplace integral equation formulation.
In the second part, the relationship between the potential energy and the total
cross-section, which is nonlinear, was linearized by a sensitivity
analysis algorithm.\cite{Lemes2006}

The cross-section sensitivity matrix established has an approximate
linear relation between the potential energy and total cross-section.
In a previous study,\cite{Lemes2014} the functional derivative of the
quantum phase shift with respect to the potential energy function
was established and coupled with the variable phase equation.
For the first time, this set of coupled differential
equations was used to establish the cross-section sensitivity matrix
within exact quantum theory. This problem is ill-posed,
since sensitivity matrix elements possess values that are
close to zero. In an attempt to circumvent this problem in a
tractable way, the Tikhonov regularization procedure was used again.

The result obtained by the inverse technique was compared with recent
\textit{ab initio} potentials\cite{Varandas2010,Pachucki2006}
and exhibited excellent agreement.
The present work presents a general strategy to obtain
an accurate inverted potential, comparable in quality with high
level potential energy models and calculations.

\section{Theoretical background}

\subsection*{General formalism of the direct problem}

Chapman-Enskog theory\cite{Chapman1953,Hirschfelder1964} provides a well-established
perturbation method to calculate the viscosity coefficient $\eta$ of a non-ideal gas
from the collision integral $\Omega^{(2,2)}.$
In a first order approach, the viscosity coefficient is given by
\begin{equation}
	\eta(T)=\frac{5(\pi mk_BT)^{1/2}}{16\pi \Omega^{(2,2)}}
	\label{visco}
\end{equation}
in which $m$ is the mass of the system, and $k_B$ is Boltzmann's constant.
The collision integral of interest to determine the viscosity coefficient is set in the form
\begin{equation}
	\Omega^{(2,2)}(T)=\frac{1}{4\pi(k_BT)^{{4}}}\int_0^\infty\exp(-E/k_BT)E^{3}Q^{(2)}dE
	\label{eqO}
\end{equation}
with $T$ the temperature and $E$ the collision energy.

The transport cross-section $Q^{(2)}$ in classical statistical mechanics is directly related
to the scattering angle, but in quantum assumptions and for the Bose-Einstein statistic,
the cross-section is associated with the phase shift,\cite{Hirschfelder1964} as follows:
\begin{equation}
	Q^{(2)}(\kappa)=
	\frac{8\pi}{\kappa^2}\sum_{l=0,2,4,\ldots}^\infty\frac{(l+1)(l+2)}{(2l+3)}
	\sin^2[\delta_{l+2}(\kappa)-\delta_l(\kappa)]
	\label{eqQ}
\end{equation}
with $\kappa=2\pi\sqrt{2mE}/h$ the wave number, $h$ Planck's constant and
$l$ the angular moment.
The phase shift $\delta_l$ is obtained from Calogero's equation,
\begin{equation}
	\frac{d\delta_l(R;\kappa)}{dR} = -\frac{1}{\kappa}
	U_{\rm eff}(R)
	\sin^2[\kappa R+\delta_l(R;\kappa)]
	\label{eqC}
\end{equation}
when $R\rightarrow \infty,$ in which $U_{\rm eff}(R)=\frac{8\pi^2\mu}{h^2}E_{\rm p}(R)+\frac{l(l+1)}{R^2}$
and $R$ is the interatomic distance and $\mu$ is the system reduced mass. More details about the
Calogero equation can be found in the published works of Lemes et al.,\cite{Lemes2014} Viterbo2014 et
al.,\cite{Viterbo2014} and Braga and Murrell.\cite{Braga1984} The viscosity coefficient as a function
of the temperature is determined by solving these equations, obeying the sequence
(\ref{eqC}), (\ref{eqQ}), (\ref{eqO}) and (\ref{visco}).

\subsection*{The inverse problem strategy}

In another approach, it is desired to obtain the potential energy function from the
viscosity coefficient in the opposite direction of the direct problem, that is,
in the sequence (\ref{visco}), (\ref{eqO}), (\ref{eqQ}) and (\ref{eqC}).
The first step consists of computing the collision integral ${\bf \Omega}^{(2,2)}$
from viscosity coefficient data using equation (\ref{visco}) through
simple algebraic manipulation. The second step consists of the evaluation
of the transport cross-section  ${\bf Q}^{(2)}$ from the collision integral
${\bf \Omega}^{(2,2)},$ and these quantities are related by equation (\ref{eqO}).
The bold style was used to represent functions in matrix form. Using the trapezoidal
quadrature, equation (\ref{eqO}) is transformed to an algebraic form
${\bf \Omega}^{(2,2)} = {\bf K}{\bf Q}^{(2)},$ the solution of which is determined
by the Tikhonov regularization method as
\begin{equation}
	{\bf Q}^{(2)}=({\bf K}^{T}{\bf K}+\lambda{\bf I})^{-1}({\bf K}^{T}{\bf\Omega}^{(2,2)}
	+\lambda {\bf Q}_0^{(2)}),
	\label{tikQ}
\end{equation}
in which ${\bf Q}_0^{(2)}$ is a first approximation for ${\bf Q}^{(2)}.$ With ${\bf Q}^{(2)}$
found, it is possible to obtain the potential energy $E_{\rm p}(R)$ from
the cross-section through equations (\ref{eqQ}) and (\ref{eqC}).

The main difficulty of this strategy is to obtain the $E_{\rm p}(R)$
value from $Q^{(2)}(\kappa),$ due to the nonlinear relationship between these two quantities.
This difficulty can be avoided by linearization of the problem as follows:
\begin{equation}
	\Delta{\bf  Q}^{(2)}={\bf S}_Q^\prime \Delta{\bf E}_p
	\label{eqlinear}
\end{equation}
in which the matrix ${\bf S}_Q^\prime$ is the sensitivity matrix, with elements
$S^\prime_Q(\kappa,R^*) = \frac{\partial Q^{(2)}(\kappa)}{\partial E_{\rm p}(R^*)}.$
Nevertheless, it is more appropriate to use equation (\ref{eqlinear}) as
\begin{equation}
	\label{lnQ}
	\Delta\ln {\bf Q}^{(2)}={\bf S}_Q  \Delta\ln {\bf E}_{\rm p}
\end{equation}
with elements of ${\bf S}_Q$ in a normalized form, such as
$S_Q(\kappa,R^*)=S^\prime_Q(\kappa,R^*)\frac{E_{\rm p}(R^*)}{Q^{(2)}(\kappa)}.$
Here, the adequate solution can not be found by an orthodox inverse matrix algorithm,
because the matrix ${\bf S}_Q$ is ill conditioned.
The inverse matrix will be given again by another Tikhonov regularization
\begin{equation}
	{\bf S}_Q^{-1} =({\bf S}_Q^T{\bf S}_Q+\lambda{\bf I})^{-1}{\bf S}_Q^T
\end{equation}
in which $\lambda$ is a parameter of regularization and is the identity matrix.
Finally, the desired solution is given by
\begin{equation}
	{\bf E}_{\rm p}^{(1)}={\bf E}_{\rm p}^{(0)} (1+({\bf S}_Q^T{\bf S}_Q+
	\lambda{\bf I})^{-1}{\bf S}_Q^T \Delta\ln {\bf Q}^{(2)})
	\label{tikEp}
\end{equation}
in which ${\bf E}_{\rm p}^{(0)}$ is an initial guess for ${\bf E}_{\rm p}.$
In this case, the regularization parameter is chosen by an L-curve, and provides
an equilibrium between a residual norm and a solution norm.

\subsection*{The alternative sensitivity matrix calculation}

The success of the present method relies on identifying a way to determine the sensitivity
matrix. In a previous work,\cite{Lemes2014} a differential equation for the functional derivative
of the quantum phase shift with respect to the potential energy function,
$S_{\delta_l}(k,R^*)=\frac{\partial\delta_l(k)}{\partial V(R^*)}$,
is established and coupled to Calogero's equation (\ref{eqC}),
\begin{equation}
	\begin{array}{clcr}
		\frac{d \delta_l}{dR}=-\frac{U_{\rm eff}(R)}{\kappa}\sin^2(\kappa R+\delta_l) \\
		\frac{ dS_{\delta_l}}{dR}=-\frac{1}{\kappa}[G(R,{R}^*)\sin^2(\kappa R+\delta_l)+2U_{\rm
		eff}(R)\sin( \kappa R+\delta_l)\cos( \kappa R+\delta_l)S_{\delta_l}]
	\end{array}
	\label{eqSp}
\end{equation}
in which $G(R,{R}^*)$ is given by
\begin{equation}
	G(R,R^*)=\frac{\partial U_{\rm eff}(R)}{\partial E_{\rm p}(R^*)}=\left\{
	\begin{array}{c}
		\frac{8\pi^2\mu}{\hbar^2},\hspace{.3cm} R = R^* \\
		0,\hspace{.3cm} R\neq R^*                       \\
	\end{array}
	\right..
	\label{eqG}
\end{equation}
Therefore, the value of $S_{\delta_l}$ at fixed values of $\kappa$ and $R^*,$ can be determined
from coupled equations (\ref{eqSp}), at $R\rightarrow \infty$ for different angular moments.
More details about this approach are given in the work of Lemes et al.\cite{Lemes2014}

By deriving equation (\ref{eqQ}) with respect to the potential energy function at
$R^*,$ the sensitivity cross-section $S_Q^\prime(k,R^*)$ will be given by

\begin{equation}\label{SenseQ}
	S_Q^\prime(\kappa,R^*)=\frac{8\pi}{\kappa^2}\sum_{l=0,2,4,\ldots}^\infty
	\frac{(l+1)(l+2)}{(2l+3)}\sin[2(\delta_{l+2}(\kappa)-\delta_l(\kappa))]\\
	\times
	\left[S_{\delta_{l+2}}(\kappa,R^*)-S_{\delta_l}(\kappa,R^*)\right]
\end{equation}
where all of the necessary information is available from the previous step. This method
provides a new, simple and exact way to establish the sensitivity cross-section matrix
${\bf S}_Q$ english in equation (\ref{tikEp}).

Finally, the sensitivity matrix for the viscosity coefficient can be obtained by

\begin{equation}
	S_\eta(T,R^*)=-\frac{5(\pi mk_BT)^{1/2}}{16\pi}\frac{1}{(\Omega^{(2,2)})^2}S_{\Omega}
\label{SenseEta}
\end{equation}
in which
\begin{equation}
	S_{\Omega}(T,R^*)=\frac{1}{4\pi(k_BT)^{{4}}}\int_0^\infty\exp(-E/k_BT)E^{3}S_Q^\prime(E,R^*)dE.
\end{equation}

There is no need to determine matrix (\ref{SenseEta}) in our method, but this matrix suggests a new
procedure to recover the potential energy function from viscosity coefficient data, such as using
$\Delta {\bf \eta}={\bf S}_{\eta}\Delta {\bf E}_p$ instead of equation
(\ref{eqlinear}). Nevertheless, preliminary studies show that the condition number of matrix
${\bf S}_\eta$  is larger than the condition number of matrix ${\bf S}_Q,$ making this
method more difficult than the procedure proposed in this study.

\section{Results and discussion}

\subsection*{Sensitivity analysis and experimental data}

\begin{figure}[ht]
	\centering
	\subfigure[]{\includegraphics[width = 8cm]{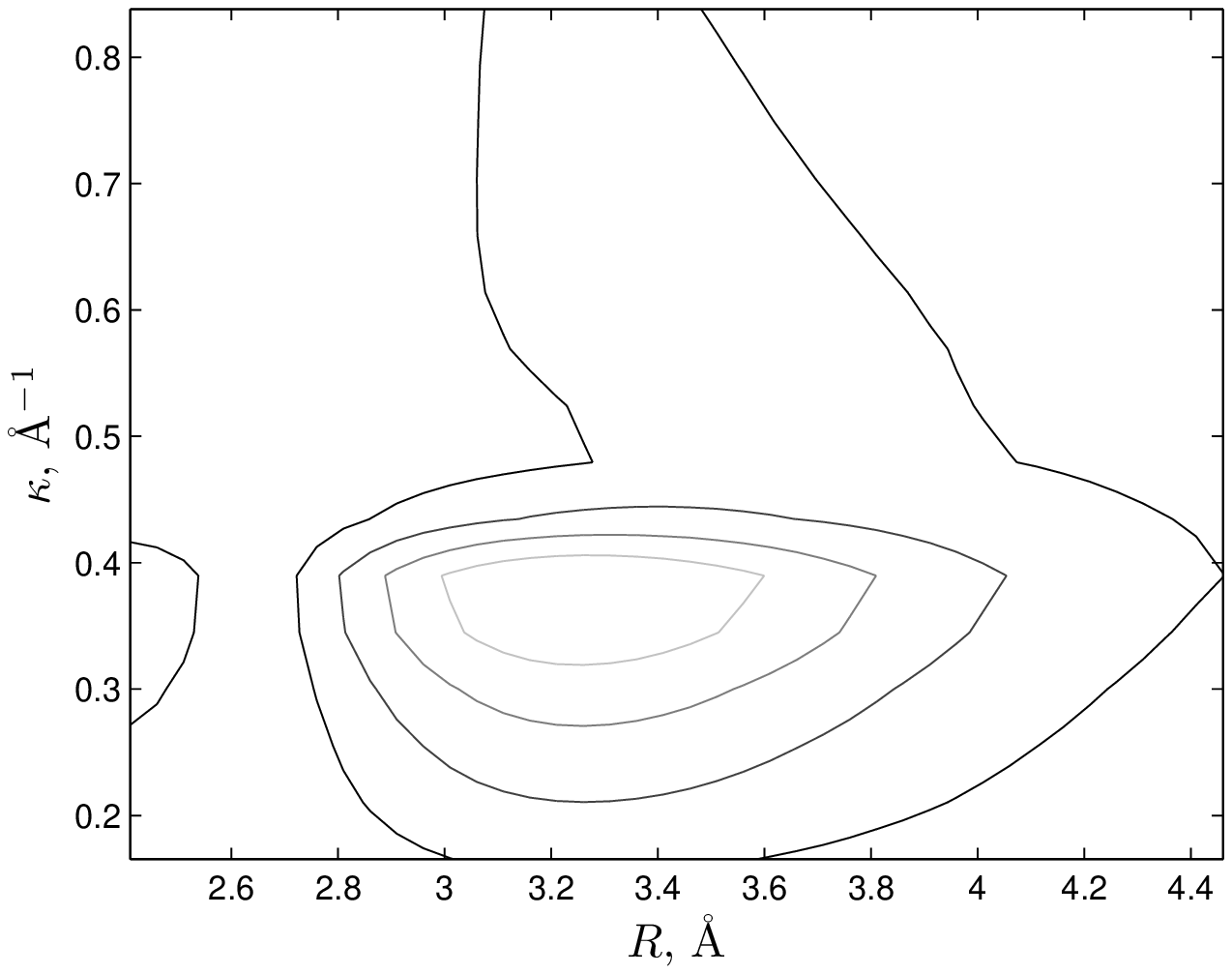}}
	\qquad
	\subfigure[]{\includegraphics[width = 8cm]{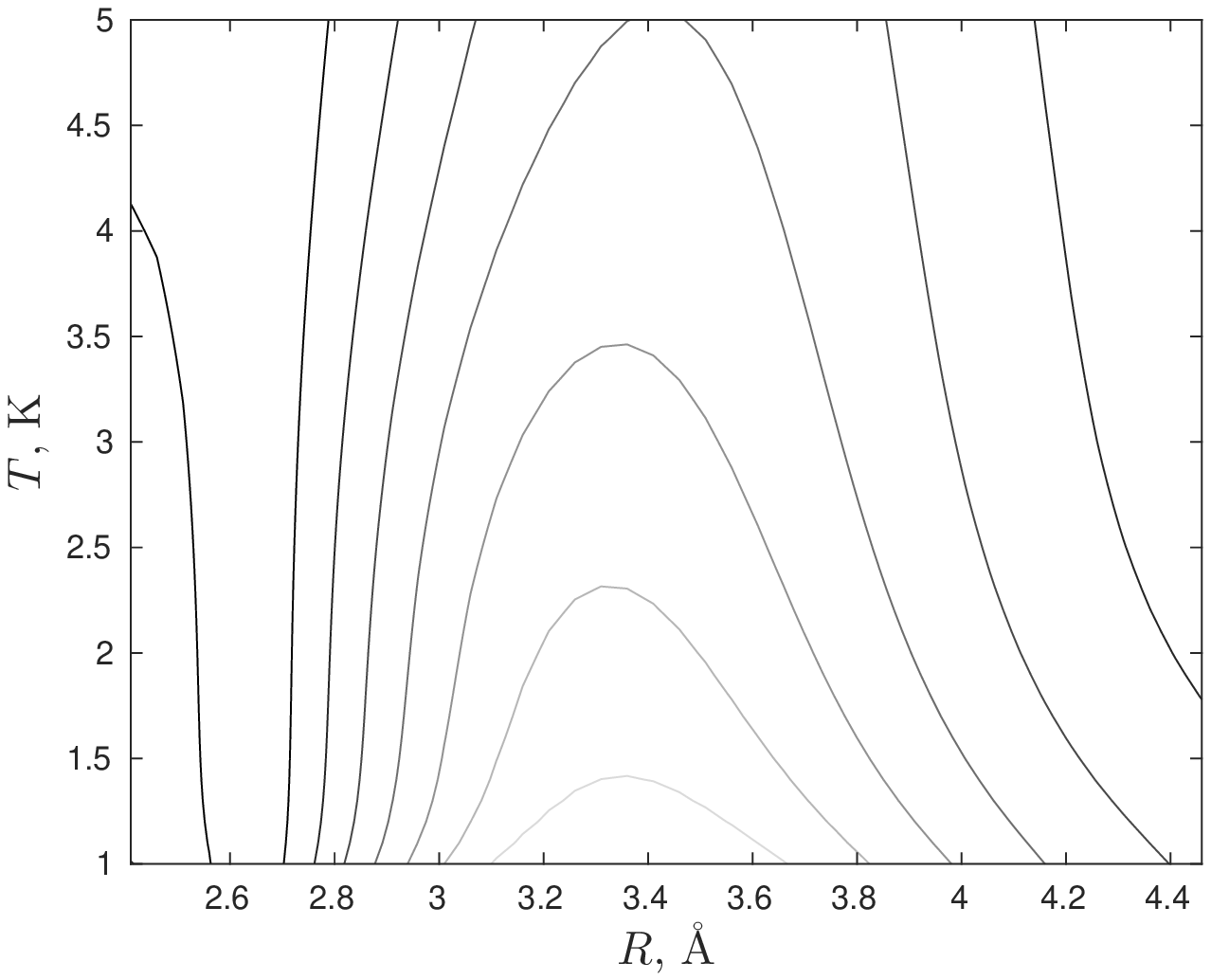}}
	\caption{Level curves: (a) for the normalized cross-section sensitivity matrix and (b) for
the normalized viscosity sensitivity matrix.}
	\label{fig1}
\end{figure}

Figure \ref{fig1}(a) shows the contour lines of the normalized sensitivity matrix to the transport
cross-section data, ${\bf S}_Q.$ This result provides important insight into the inverse
procedure, such as the choice of the experimental data range to be used.

\begin{center}
	{\bf Figure 1}
\end{center}

From Figure \ref{fig1}(a), it can be observed that the transport cross-section with
$\kappa$ between 0.2 and 0.5~$\text{\AA}^{-1}$ has a larger value
of sensitivity at interatomic distances between 3 and 4~{\AA}. This region of $\kappa$
provides experimental data that are more adequate to obtain the long range potential
between 3 and 4~{\AA} because small changes in the potential can cause large changes in
the transport cross-section. Since the $\kappa$ values of approximately 0.1~$\text{\AA}^{-1}$
correspond to energy values of approximately 0.1~K, it is better to use viscosity
experimental data obtained at low temperatures for the inverse procedure.
This conclusion can also be obtained from Figure \ref{fig1}(b), showing the sensitivity
level curves for viscosity data.

Only a small number of papers have reported experimental data on viscosity coefficients
at low temperatures. For the temperature range required to refine the potential curve
between 3 and 4~{\AA}, that is, below 5 K, the reference of Becker et al.\cite{Becker1954}
presents 11 values. However, the quality of these data has been questioned\cite{Bich2007} due
to the use of an old reference value for equipment calibration,
resulting in a positive deviation of approximately 5~\%
as estimated by Bich and Vogel.\cite{Bich2007}

\begin{figure}[ht]
	\centering
	\includegraphics[scale = 1]{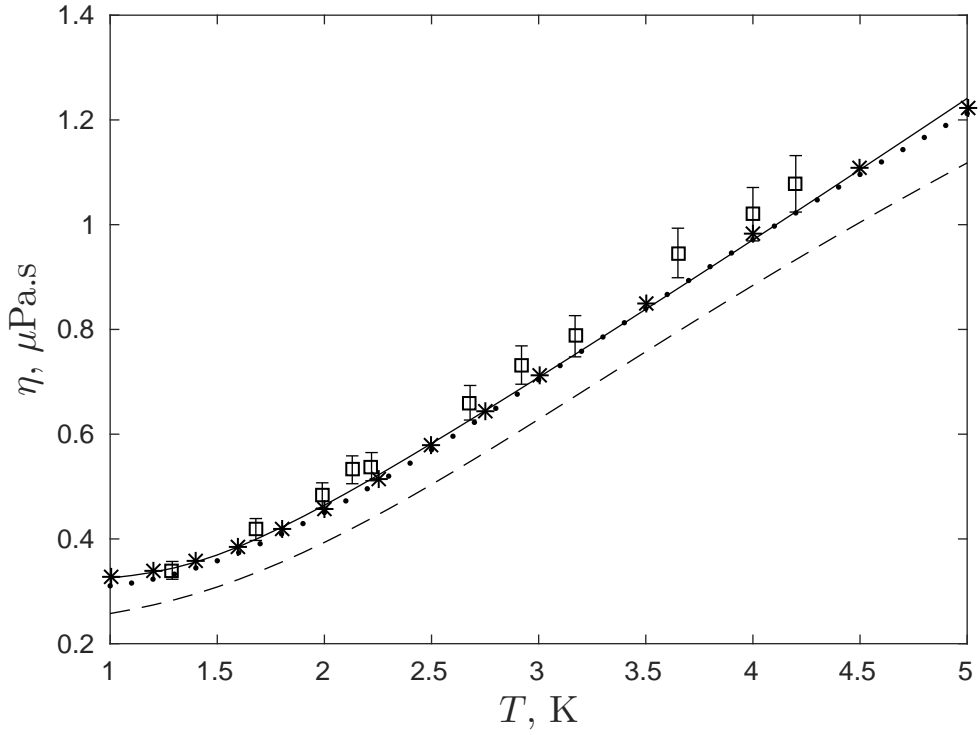}
	\caption{Figure 2: Viscosity coefficients of $^4{\rm He}:$
	(squares) the experimental data from Becker et al.,\cite{Becker1954}
	(asterisks) the calculated data from Cencek et al.,\cite{Cencek2012}
	(solid curve) the data generated with equation (\ref{Nachereq}),
	(dashed curve) the data calculated with $E_{\rm p}^{(0)}$ and
	(dotted curve) the data calculated with $E_{\rm p}^{(3)}$.}
	\label{fig2}
\end{figure}

To generate the necessary data of viscosity coefficients, the equation
\begin{equation}
	\frac{\eta}{\rm Pa.s} =
	2.113\times10^{-7}\left(T/{\rm K}\right)^{1.1}\left\{1+1.16\exp\left[-2.44\left(\log ({T}/{\rm K})
	+0.56\right)^2\right]\right\},
	\label{Nachereq}
\end{equation}
was used.\cite{Nacher1994} From this equation, 41 viscosity values,
between 1 and 5~K, were determined and used to refine the potential curve.
The values adjusted by the equation (\ref{Nachereq}),
shown in Figure \ref{fig2}, agreed with the calculated results with the
most precise potential for description of the system,\cite{Przybytek2010,Cencek2012}
with an average error of 0.8735~\%, an error lower than that reported for the
experimental data (5~\%). Therefore, henceforth we will refer to equation
(\ref{Nachereq}) as experimental data.

\begin{center}
	{\bf Figure 2}
\end{center}

\subsection*{Recent potentials and initial information}

In our inversion procedure, an initial potential curve,
$E_{\rm p}^{(0)},$  is given as \textit{a priori} information.
This approximate potential curve is given by
\begin{equation}\label{Ep0}
	E_{\rm p}^{(0)}(R) = \left\{ {\begin{array}{*{20}{l}}
		{E_{\rm p}(R),R \le {R_1}}\\
		{E_{\rm p}(R) \times \left\{ {1 + \varepsilon \left[ {1 - {{\cos }^2}\left[ {\pi \left( {\frac{{R -
			{R_1}}}{{{R_2} - {R_1}}}} \right)} \right]} \right]} \right\},{R_1} < R < {R_2}}\\
		{E_{\rm p}(R),R \ge {R_2}}
		\end{array}} \right.
\end{equation}
as suggested in the reference of Keil and Danielson,\cite{Keil1988}
in which $R_1=2.4,$ $R_2=4.5,$ and $\varepsilon=0.2$ were used to modify
the potential curve $E_{\rm p}(R)$ between 3 and 4~{\AA}. The
function $E_{\rm p}(R)$ is the \textit{ab initio} potential
curve proposed in 2010,\cite{Varandas2010} henceforth considered
to be the reference potential. The $E_{\rm p}(R)$ curve was previously
tested to evaluate the second virial coefficient between 3 and 100 K
with excellent agreement with the experimental data.\cite{Costa2013}
When this potential is compared with the most accurate potential
describing the system,\cite{Przybytek2010} the difference is
less than $22.4\;\mu{\rm eV},$ between 2.4 and 4.5~{\AA}.
Nevertheless, both curves, Varandas\cite{Varandas2010} and Przybytek et al.\cite{Przybytek2010},
describe the viscosity coefficient with an error less than 5~\% for temperatures
between 1 and 5~K. Therefore, both \textit{ab initio} potential curves,
namely, those of Varandas\cite{Varandas2010} and Przybyteke et al.\cite{Przybytek2010}, are equally
suitable to describe the viscosity coefficient between 1 and 5~K.

The difference between the potential curves $E_{\rm p}^{(0)}$ and
$E_{\rm p}$ is controlled by the parameter $\varepsilon,$ when
$\varepsilon$ is equal to 0.2 the average difference is approximately 10~\%.
In this case, if the potential curve $E_{\rm p}^{(0)}$ is used to calculate
the viscosity data the average error found was 12.46~\%, which is greater than
the experimental error of 5~\%. Therefore, the calculated viscosity data from the potential curve
$E_{\rm p}^{(0)}$ are not within the experimental error,
and it is not suitable for estimating the viscosisty coefficient.

The determination of the viscosity coefficient from a potential energy curve involves three
steps: the determination of the phase shift using equation (\ref{eqC});
followed by the determination of the transport cross-section using equation (\ref{eqQ});
and finally the determination of the collision integral using equation (\ref{eqO}) from which
the viscosity can be readily calculated. Numerical integration of Calogero's equation (\ref{eqC})
was performed using the Euler algorithm, with a step of $10^{-2}$~{\AA} between 1.5 and 70000~{\AA}.
Next, in equation (\ref{eqQ}), the sum is carried over even angular
moments between 0 and 20. Finally, the trapezoidal rule was used to numerically
integrate equation (\ref{eqO}) over 100 points between $6.0\times10^{-3}$ and 5.4 meV.
Figure \ref{fig2} shows the experimental data\cite{Nacher1994} together with the
values calculated by equations (\ref{eqC}), (\ref{eqQ}), (\ref{eqO}) and (\ref{visco}),
using the potential function $E_{\rm p}^{(0)}.$

\subsection*{The inverse procedure}

\begin{figure}[ht]
	\centering
	\includegraphics[scale = 1]{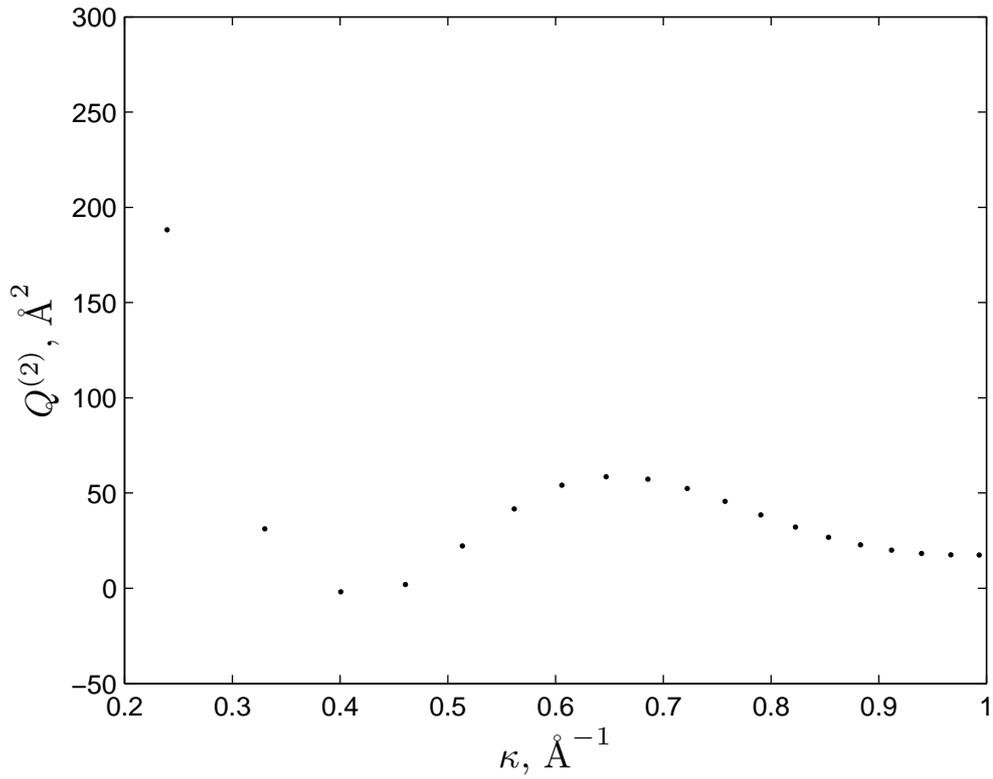}
	\caption{Inverted cross section.}
	\label{fig3}
\end{figure}

The inverse problem consists of the determination of $E_{\rm p}(R)$
using known viscosity data, $\eta(T).$ Therefore, one has to solve
the set of four equations in the reverse sequence:
(\ref{visco}), (\ref{eqO}), (\ref{eqQ}) and (\ref{eqC}).
The first step provides the values of the transport cross-section,
which are shown in Figure \ref{fig3}. This step involves equation
(\ref{eqC}), which is a linear ill-posed problem, whose solution
is determined by the Tikhonov regularization method,
shown in equation (\ref{tikQ}).

\begin{center}
      {\bf Figure 3}
\end{center}

The result shown in Figure \ref{fig3} was obtained with
$\lambda=2.5\times 10^{-5}$ and using a matrix \textbf{K} with
dimensions of 41 (temperature) $\times$ 101 (energy). The value of
${\bf Q}_0^{(2)}$ is obtained from $E_{\rm p}^{(0)}$
following the direct method.

In the second part of the inverse procedure, the equations (\ref{eqQ}) and (\ref{eqC})
are rewritten as (\ref{lnQ}), which is another linear ill-posed problem,
the solution of which is provided by equation (\ref{tikEp}). Before using equation (\ref{tikEp})
to refine the potential energy function, one needs to determine the matrix \textbf{S}
from equations (\ref{eqSp}) and (\ref{SenseQ}). The coupled differential
equations (\ref{eqSp}) were solved using Euler's method with initial conditions
$\delta_l(R_0)=-\kappa R_0,$ $S_{\delta_l}(R_0)=0$ and $R_0 = 1.5$~{\AA}.
In the limit $R\rightarrow \infty,$ $S_{\delta_l}$ is the desired sensitivity value
at fixed $\kappa,$ $R^*$ and $l.$
Therefore, solving the coupled differential equations (\ref{tikEp}), for different values of
$R^*,$ $\kappa$ and $l$, we obtain the ${S}_{\delta_l}(\kappa, R*)$ values from which,
using equation (\ref{eqG}), we calculated ${S}_Q(\kappa, R^*)$ and consequently ${\bf S}.$

Now, knowing ${\bf S},$ equation (\ref{tikEp}) can be used in the following manner:
first, an initial guess, ${\bf E}_{\rm p}^{(0)},$ is given for the unknown potential,
and the value of ${\bf Q}_0^{(2)}$ is computed for this first potential estimation.
The difference, $\Delta\ln {\bf Q}^{(2)},$ is then calculated between the
$\ln{\bf  Q}^{(2)}$ obtained from the first part and $\ln {\bf Q}_0^{(2)}.$
A correction to the initial potential energy function can be evaluated interactively by
using equation (\ref{tikEp}). The refined potential ${\bf E}_{\rm p}^{(1)}$
minimizes the Tikhonov criterion function, for which the balance between the residual norm
and the solution norm is given by $\lambda = 19.$

\begin{figure}[ht]
	\centering
	\includegraphics[scale = 1]{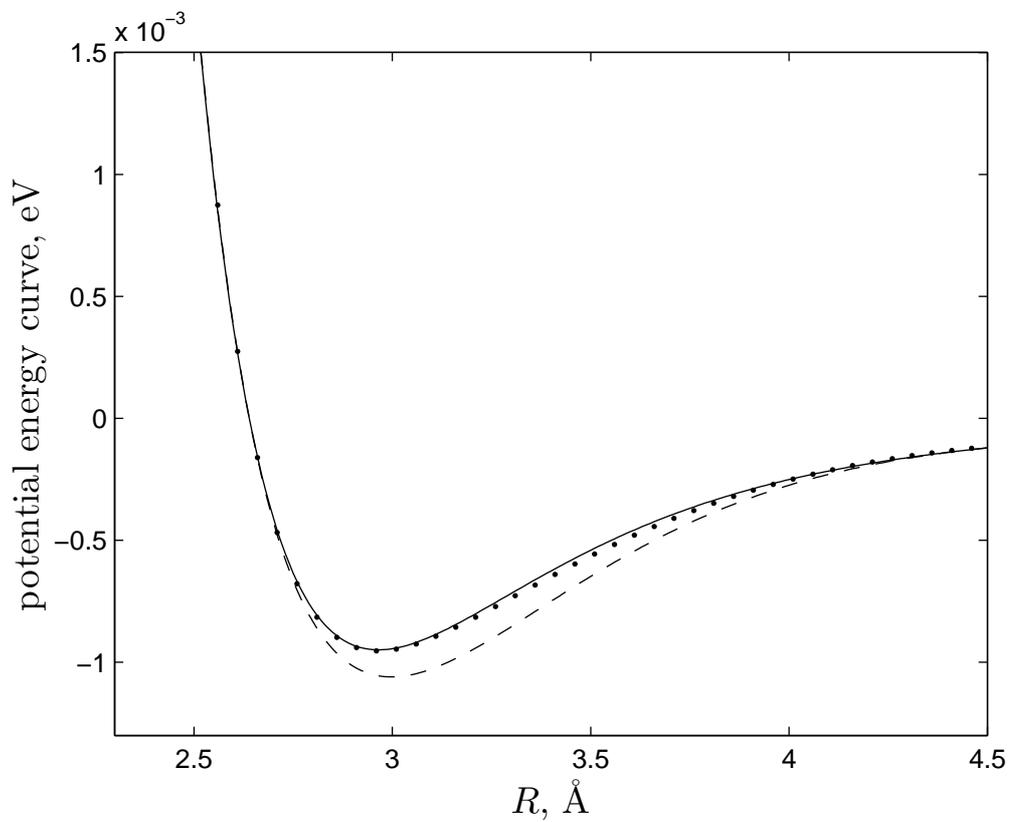}
	\caption{Interatomic potentials:
	(solid curve) the reference potential energy function,
	(dashed curve) the initial guess and (dotted curve) from this work.}
	\label{fig4}
\end{figure}

The potential curve $E_{\rm p}^{(3)},$ obtained after three iterations,
is shown in Figure \ref{fig4} together with the reference potential $E_{\rm p}$
and the initial choice $E_{\rm p}^{(0)}.$ The refined potential $E_{\rm p}^{(3)}$
has an average error of 2.6~\% between 2.4 and 4.5~{\AA} compared with the
\textit{ab initio} potential curve.\cite{Varandas2010}
The potential depth is -11.1~K for the potential obtained in this work,
while for the reference potential, the depth is -11.0~K;
therefore, an error of less than 1~\% is observed.
The equilibrium distance is 2.96~{\AA} for both potentials

\begin{center}
	{\bf Figure 4}
\end{center}

The viscosity coefficient was determined using refined potential $E_{\rm p}^{(3)},$
this result together with experimental data are shown in Figure \ref{fig2},
for temperatures between 1 and 5~K. The improved potential is adequate to describe
the experimental data with an average error of 1.6823~\% against 12.4632~\% when
$E_{\rm p}^{(0)}$ was used. The average error found with the $E_{\rm p}^{(3)}$
potential is less than the experimental error (5~\%), therefore the inverted results
provided a better potential than the initial curve $E_{\rm p}^{(0)}.$ The potential
$E_{\rm p}^{(3)},$ obtained directly from the experimental data, had equivalent results
when compared with the \textit{ab initio} potential presented in reference,\cite{Varandas2010}
in which the average error is 1.1173~\%. Both the refined potential $E_{\rm p}^{(3)}$
and the \textit{ab initio} potential $E_{\rm p}$\cite{Varandas2010} are in agreement
with the experimental value, with errors less than the experimental errors.
Therefore, both potential curves, $E_{\rm p}^{(3)}$ and $E_{\rm p},$ are equally
suitable to describe the viscosity coefficient between 1 and 5~K.

\section{Conclusions}

The long range potential function for helium was determined in the present work
from viscosity coefficient data at temperatures below 5~K.
To the best of our knowledge, this approach has never been reported. In
this case, quantum assumptions and the Bose-Einstein statistic are required
to connect viscosity coefficient with interatomic potential. This relationship
is nonlinear, and a common method to solve this inverse problem cannot be
employed. To circumvent this difficulty, the sensitivity matrix was used
to linearize this problem. The initial problem was rewritten as a linear
Fredholm integral equation of the first order, whose solution
is given by the Tikhonov method. The success of the proposed method
depends on finding a cross-section sensitivity matrix by a consistent and
robust procedure.

In a previous work,\cite{Lemes2014} a differential equation for the functional
derivative of the quantum phase shift with respect to the potential energy was
established and coupled to Calogero's equation. In the present work,
this previous result was used to establish the cross-section
sensitivity matrix within exact quantum theory, for
the first time.

Finally, the long range potential curve was obtained from experimental data
and compared with recent \textit{ab initio} potentials,\cite{Varandas2010}
showing an excellent agreement. The average difference between both is less
than 2.6~\% between 3 and 4~{\AA}, with an even smaller difference
in the potential depth, -11.1~K against -11.0~K. The potential depth is important result,
which shows that there is excellent agreement with a bound state prediction of the
$^4{\rm He}$ dimer.\cite{Costa2013} The refined potential describes the
viscosity coefficient with an average error of 1.6823~\% that is less than the experimental
error (5~\%), a result similar to that found for the \textit{ab initio} potential (1.1173~\%).

The present algorithm is general and can be used to determine the
potential energy function whenever \textit{ab initio} calculations
cannot be used. The method could be applied without difficulty
to an inverted second virial coefficient or other transport properties at
low temperatures when quantum assumptions are needed.

\section*{Acknowledgements}

We would like to thank CNPq and FAPEMIG for their financial support.

\clearpage

\section*{Figures}

\end{document}